\documentclass[prd,nofootinbib,tightenlines,showpacs]{article}
\def\journal#1, #2, #3#4, #5#6#7#8    {
    {#1~} {#2}  (#5#6#7#8) #3#4}

\usepackage{amssymb}
\usepackage{amsmath}
\usepackage{lscape}
\begin{document}


\renewcommand{\thesection}{\arabic{section}}
\renewcommand{\thesubsection}{\thesection.\arabic{subsection}}
\renewcommand{\theequation}{\arabic{equation}}
\newcommand{\pv}[1]{{-  \hspace {-4.0mm} #1}}

\baselineskip=14pt


\begin{center}
{\bf  \Large  Electrodynamics on $\kappa$-Minkowski space-time}
 
\bigskip

 E. Harikumar {\footnote{harisp@uohyd.ernet.in}} \\  
School of Physics, University of Hyderabad, Central University P O, Hyderabad, AP, PIN 500046, India \\[3mm]

 T. Juri\'c 
 {\footnote{e-mail: tjuric@irb.hr}}, S. Meljanac {\footnote{e-mail: meljanac@irb.hr}},
 \\  
 Rudjer Bo\v{s}kovi\'c Institute, Bijeni\v cka  c.54, HR-10002 Zagreb,
Croatia \\[3mm]


\end{center}
\setcounter{page}{1}
\bigskip





\begin{abstract} 
In this paper, we derive Lorentz force and Maxwell's equations on kappa-Minkowski space-time up to the first order in the deformation parameter. This is done by elevating the principle of minimal coupling to non-commutative 
space-time. We also show the equivalence of minimal coupling prescription and Feynman's approach.  It is shown that the motion in kappa space-time  can be interpreted as motion in a background gravitational field, which is induced by this non-commutativity.  In the static limit, the effect of kappa deformation is to scale the electric charge. We also show that the laws of electrodynamics depend on the mass of the charged particle, in kappa space-time. 

\end{abstract}
 




\section{Introduction}

Quantum gravity effects are expected to lead to space-time uncertainties \cite{Dopl} at Planck scale. Non-commutative geometry provides a natural way to incorporate this microscopic structure of space-time. $\kappa$-deformed space-time \cite{JL1,JL2} is a prototype of the Lie algebraic type noncommutative spacetime, fuzzy sphere being the well known example of this family \cite{madore,balbook}. The $\kappa$-deformed space-time is known to emerge naturally in the low energy limit of certain quantum gravity models. It is also the space-time associated with doubly special relativity \cite{KG,AC,Ghosh}. In recent years, algebraic structure and symmetries of $\kappa$-space-time have been investigated in detail \cite{md,sm,sm1}. 

Generically field theory models on non-commutative space-time do have highly non-local and non-linear interactions, and are characterized by an interdependence of high and low energy behaviour, known as UV/IR mixing, which had been studied in detail in field theory models on Moyal spacetime as well as on $\kappa$ spacetime \cite{uvir}. In non-commutative space-times, Lorentz symmetry, in the usual sense is broken, but it is shown that this symmetry can be retained by a Hopf algebra approach \cite{chai}. Thus, the conventional notions of field quanta can be generalized to non-commutative field theories. Following these developments, field theory models on $\kappa$-deformed space-time have been constructed and studied \cite{JL3,sm2}. Investigations, trying to obtain bounds on $\kappa$ deformation parameter using experimental and observational results are being carried out in last couple of years \cite{bol,ab,msk,akk}.

Construction and study of $U(1)$ gauge theory on $\kappa$-space-time using star product approach was taken up in \cite{jw1}. Using Feynman's approach, Maxwell's equations on $\kappa$-space-time was obtained in \cite{eh}.

In Feynman's approach, starting with Newton's equation of motion and assumed (quantum) commutators between coordinates and velocities, one derives the homogeneous Maxwell's equations by the repeated application of Jacobi identities \cite{Dyson}. This method has been generalized to relativistic case in \cite{Tanimura}. It has also been shown that the quantum mechanical particle consistently interact with scalar, gauge, and gravitational fields only. In the commutative space-time, it is known that the results obtained by Feynman's approach and minimal prescription are equivalent \cite{Dyson,minfeyn}. 
Feynman's approach has been generalized to obtain inhomogeneous Maxwell's equation in \cite{abyg} and various other aspects of this method has been studied in \cite{abyg1}. In recent times, this method has been used to obtain Maxwell's equation in Moyal space time also \cite{ncfey}. In \cite{minfeyn}, it was shown that by assuming the minimal coupling of gauge field and the ensuing relation between kinetic and conjugate momenta, one can derive Lorentz force equation and Maxwell's equations. In this way, one can work with Poisson brackets rather than (quantum) commutators in the Feynman's approach. Thus, this approach of \cite{minfeyn} allows to take a classical limit of the obtained equations, in a proper fashion.

In this paper, we generalize a variant of Feynman's approach \cite{minfeyn} to $\kappa$-space-time and derive the deformed Maxwell's equations and force equation valid up to 1st order in deformation parameter. In our approach, we do power series expansion of the non-commutative coordinates, momenta, as well as functions of non-commutative coordinates and momenta in terms of commutative coordinates, momenta and deformation parameter (keeping terms up to the 1st order in deformation parameter). We also exploit the generalization of minimal coupling prescription to $\kappa$-space-time in our calculations.

This paper is organized as follows. In the next section, we briefly recall the Feynman's approach using minimal prescription \cite{minfeyn}. Here we show, how the force equation as well as all the Maxwell's equations can be derived. Our main results are presented in section 3. In subsection 3.1, we discuss the derivation of force equation on $\kappa$ space-time for a electrically neutral particle. The force equation we obtain here is valid up to first order in the deformation parameter $a$.  In subsection 3.2, we derive the Lorentz force equation and in subsection 3.3,  Maxwell's equation for a charged particle in $\kappa$-space-time is obtained.  Here also, all equations derived are valid up to first order in the deformation parameter $a$. In subsection 3.4, we discuss the natural realization of the coordinates of $\kappa$-space-time and obtain the Maxwell's equation in this realization. Finally we conclude with discussion in section 4.

We work with $\eta_{\mu\nu}=(+,-,-,-)$.


\section{Minimal Coupling and Feynman's approach to Electrodynamics}

We start with the same   basic assumptions as those in Feynman's approach \cite{Dyson, Tanimura,minfeyn}.  The coordinates of a relativistic particle in 4-D Minkowski space-time is described by $x_{\mu}(\tau),(\mu=0,1,2,3)$, where $\tau$ is a parameter, and they satisfy the following commutation relations
\begin{equation}
[x_{\mu}(\tau),x_{\nu}(\tau)]=0, \qquad [x_{\mu}(\tau),\dot{x}_{\nu}(\tau)]=-\frac{i}{m} \eta_{\mu\nu},
\end{equation}
where $\dot{x}_{\mu}=\frac{d x_{\mu}}{d \tau}$. Newton's equation is also assumed
 \begin{equation}
F_{\mu}(x,\dot{x})=m\ddot{x}_{\mu}.
\end{equation}
We introduce kinetic momentum $\pi_{\mu}=m\dot{x}_{\mu}$, so that we have
\begin{equation}\label{com1}
[x_{\mu},\pi_{\nu}]=-i \eta_{\mu\nu},
\end{equation}
and we can write $\pi_{\mu}(\tau)$ explicitly as 
\begin{equation}\label{kinetic}
\pi_{\mu}=p_{\mu}-eA_{\mu}(x),
\end{equation}
 where $eA_{\mu}(x)$ is, for now, arbitrary function of $x$, and $p_{\mu}(\tau)$ is canonical
 momentum satisfying 
\begin{equation}\label{canonical}
[p_{\mu},p_{\nu}]=0, \qquad [x_{\mu},p_{\nu}]=-i \eta_{\mu\nu}
\end{equation}
This is the principal of minimal coupling \cite{minfeyn}.
It is obvious that with $F_{\mu}=\frac{d \pi_{\mu}}{d \tau}$, and taking the derivative with respect to $\tau$ of Eqn.(\ref{com1}) one gets the following relations
\begin{equation}\label{com.sila}
[x_{\mu},F_{\nu}]=-\frac{1}{m}[\pi_{\mu},\pi_{\nu}], \qquad [\pi_{\mu},\pi_{\nu}]=-ieF_{\mu\nu}(x),
\end{equation}
where $F_{\mu\nu}=\partial_{\mu}A_{\nu}-\partial_{\nu}A_{\mu}$. Because of (\ref{canonical}) we have useful identities
\begin{equation}\label{identity}
[x_{\mu},f(x,p)]=-i\frac{\partial f}{\partial p^{\mu}},
\qquad
[p_{\mu},f(x,p)]=i\frac{\partial f}{\partial x^{\mu}}.
\end{equation}
Here (and from now on) we take the ordering prescription where $x$ is put always on the left, and $p$ on the right.
Force $F_{\mu}(x,\dot{x})$ can be understood as a function of $x$ and $p$, that is $F_{\mu}(x,p)$, and this is important because we only know how to integrate over commuting variables. So, using (\ref{identity}) we can now integrate (\ref{com.sila}) over $p^{\mu}$ and get
\begin{equation}
F_{\nu}=\frac{e}{m}F_{\nu\mu}p^{\mu}+\widetilde{G}_{\nu}(x),
\end{equation}
where $\widetilde{G}_{\nu}(x)$ is a function of $x$ and we have use the prescription that $p$ goes to the right (for constructing hermitian operators we can symmetrize  $xp\rightarrow \frac{1}{2}(xp+px)$ ). Now, using the definition of kinetic momentum $\pi_{\mu}=m\dot{x}_{\mu}$, relation (\ref{kinetic}), and defining $G_{\mu}(x)=\widetilde{G}_{\mu}(x)+\frac{e^{2}}{m}F_{\mu\nu}A^{\nu}$  we get the Lorentz force
\begin{equation}\label{Lorentz}
F_{\mu}=G_{\mu}(x)+eF_{\mu\nu}\dot{x}^{\mu}.
\end{equation}
In the minimal coupling approach all the Jacobi identities are  satisfied by construction, i.e.,
\begin{equation}\begin{split} \label{allJacobi}
[x_{\mu},[x_{\nu},x_{\rho}]]+[x_{\nu},[x_{\rho},x_{\mu}]]+[x_{\rho},[x_{\mu},x_{\nu}]]&=0, \\
[x_{\mu},[x_{\nu},\pi_{\rho}]]+[x_{\nu},[\pi_{\rho},x_{\mu}]]+[\pi_{\rho},[x_{\mu},x_{\nu}]]&=0,\\
[x_{\mu},[\pi_{\nu},\pi_{\rho}]]+[\pi_{\nu},[\pi_{\rho},x_{\mu}]]+[\pi_{\rho},[x_{\mu},\pi_{\nu}]]&=0,\\
[\pi_{\mu},[\pi_{\nu},\pi_{\rho}]]+[\pi_{\nu},[\pi_{\rho},\pi_{\mu}]]+[\pi_{\rho},[\pi_{\mu},\pi_{\nu}]]&=0\\
\end{split}\end{equation}
The first two Eqn. in (\ref{allJacobi}) are trivially satisfied, the third is in Feynman's approach equal to the statement that $F_{\mu\nu}$ is a function of $x$ and the fourth yields homogeneous Maxwell Eqn.
\begin{equation}
\partial _{\mu}F_{\nu\rho}+\partial _{\nu}F_{\rho\mu}+\partial _{\rho}F_{\mu\nu}=0.
\end{equation}
If we take (\ref{Lorentz}) as a definition of $G_{\mu}(x)$ it is straight forward to see

\begin{equation}
[\pi_{\mu},G_{\nu}]-[\pi_{\nu},G_{\mu}]=0, \qquad \partial_{\mu}G_{\nu}-\partial_{\nu}G_{\mu}=0,
\end{equation}
which means that $G_{\mu}=\partial_{\mu}\phi(x)$. We can conclude that minimal coupling and Feynman's approach are in complete correspondence. From the definition of the commutator we can show that
\begin{equation}
[\pi_{\nu},[\pi_{\mu},[\pi^{\mu},\pi^{\nu}]]]=0,
\end{equation}
and by defining 
\begin{equation}\label{inhomo}
[\pi_{\mu},[\pi^{\mu},\pi^{\nu}]]=ej^{\nu},
\end{equation}
we get
\begin{equation}\label{current}
[\pi_{\nu},j^{\nu}]=0, \qquad \partial_{\mu}j^{\mu}=0.
\end{equation}
Thus we see that $j_{\mu}(x)$ is the conserved current, and by definition Eqn.(\ref{inhomo}) gives the inhomogeneous Maxwell's equations.
\begin{equation}
\partial_{\mu}F^{\mu\nu}=j^{\nu}.
\end{equation}
Now we have the complete set of Maxwell equations which are covariant and we recognize $A_{\mu}(x)$ as a gauge field, and $e$ as an electric charge of a particle.


\section{$\kappa$-deformed Electrodynamics}

\subsection{$e=0$ case}
Minimal coupling approach seems natural for exploring non-commutative spaces, all it takes is to substitute $x_{\mu}\rightarrow\hat{x}_{\mu}$, where $[\hat{x}_{\mu},\hat{x}_{\nu}]\neq0$. We will consider a class of non-commuting spaces, the so called $\kappa$-Minkowski space-time, which are defined by
\begin{equation}\label{kappaM}
[\hat{x}_{\mu},\hat{x}_{\nu}]=i(a_{\mu}\hat{x}_{\nu}-a_{\nu}\hat{x}_{\mu}),
\end{equation}
where $a_{\mu}$ is the deformation parameter and $\hat{x}_{\mu}$ is a non-commuting coordinate operator. In the case when $a\rightarrow0$, we have $[\hat{x}_{\mu},\hat{x}_{\nu}]\rightarrow0$, that is $\hat{x}_{\mu}\rightarrow x_{\mu}$, so we take a perturbative approach to find the realization of $\hat{x}_{\mu}$ in terms of operators $x_{\mu}$ and $p_{\mu}$ from the commutating space, up to the first order in the deformation parameter $a_{\mu}$. So we write 
\begin{equation}\label{xhat}
\hat{x}_{\mu}=x_{\mu}+\delta\hat{x}_{\mu}(a),
\end{equation}
where $\delta\hat{x}_{\mu}(a)$ can be constructed from $x_{\mu}$, $p_{\mu}$ and $a_{\mu}$ as
\begin{equation}\label{varx}
\delta\hat{x}_{\mu}(a)=\alpha x_{\mu}(a\cdot p)+\beta(x\cdot a)p_{\mu}+\gamma(x\cdot p)a_{\mu}, \qquad \alpha,\beta,\gamma \in \mathbb{R}
\end{equation}
 Taking into account that (\ref{kappaM}) must be satisfied up to the first order in deformation parameter $a_{\mu}$ we get the constraint on the real parameters $\alpha,\beta,$ and $\gamma$
\begin{equation}
\gamma-\alpha=1, \qquad \beta \in \mathbb{R}.
\end{equation}
Now we have to construct the non-commutative momentum operator $\hat{p}$, but we are missing the relation $[\hat{p}_{\mu},\hat{x}_{\nu}]=?$, all we know that in the zeroth order in $a_{\mu}$ Eqn. (\ref{canonical}) holds. First let us consider the $e=0$ case and postulate $\dot{\hat{x}}_{\mu}(e=0)\equiv\frac{1}{m}\hat{p}_{\mu}$, then taking the derivative of (\ref{kappaM}) with respect to $ \tau$ gives 
\begin{equation}\label{anti}
[\hat{p}_{\mu},\hat{x}_{\nu}]+[\hat{x}_{\mu},\hat{p}_{\nu}]=i(a_{\mu}\hat{p}_{\nu}-a_{\nu}\hat{p}_{\mu}),
\end{equation}
 which only fixes the antisymmetric part of $[\hat{p}_{\mu},\hat{x}_{\nu}]$. We can take $\hat{p}_{\mu}$ to be 
 \begin{equation}
 \hat{p}_{\mu}=p_{\mu}+\delta\hat{p}_{\mu}(a),
 \end{equation}
 and demand that (\ref{anti}) and Jacobi identities between $\hat{p}_{\mu},\hat{x}_{\nu}$ and $\hat{x}_{\rho}$ must be satisfied up to the first order in $a$ and get the explicit form of $\delta\hat{p}_{\mu}(a)$. This construction is equivalent to just taking the form of $\delta\hat{x}_{\mu}(a)$ given in (\ref{varx}) and substitute $x$ with $p$ (which is also equivalent with $\dot{p}_{\mu}=0$) and then we get
 \begin{equation}\label{p-nonc}
 \hat{p}_{\mu}=p_{\mu}+(\alpha+\beta)(a\cdot p)p_{\mu}+\gamma a_{\mu}p^{2}.
 \end{equation}
 Now we have
 \begin{equation}\label{p}
 [\hat{p}_{\mu},\hat{p}_{\nu}]=0,
 \end{equation}
 Using (\ref{p-nonc}) we have 
 \begin{equation}\label{comm-NC}
 [\hat{p}_{\mu},\hat{x}_{\nu}]=i\eta_{\mu\nu}(1+s(a\cdot p)) + i(s+2)a_{\mu}p_{\nu}+i(s+1)a_{\nu}p_{\mu}, \qquad s=2\alpha + \beta.
 \end{equation}
 We are considering $e=0$ case, so there is no difference between canonical and kinetic momentum. Analogous  to commutative space we have the Newton-like equation
 \begin{equation}
 \hat{F}_{\mu}(e=0)\equiv\hat{G}_{\mu}=\frac{d \hat{p}_{\mu}}{d\tau}.
 \end{equation}
 Taking the derivative with respect to $\tau$ of Eqn. (\ref{comm-NC}) we get
 \begin{equation}\label{com-F}
 [\hat{G}_{\mu},\hat{x}_{\nu}]=i\eta_{\mu\nu}s(a\cdot G)+i(s+2)a_{\mu}G_{\nu}+i(s+1)a_{\nu}G_{\mu},
 \end{equation}
 where we used (\ref{p}), and the fact that all equations from Section 2. hold up to the zeroth order in $a$. We want to find $\hat{G}_{\mu}$, but we can not simply integrate (\ref{com-F}). The force $\hat{G}_{\mu}$ can be written as
 \begin{equation}\label{F-e=0}
 \hat{G}_{\mu}=G_{\mu}(x)+\delta\hat{G}_{\mu}(a),
 \end{equation}
 and combining (\ref{com-F}) and (\ref{F-e=0}) we can get an equation for $\delta\hat{G}_{\mu}(a)$
 
 \begin{equation}\begin{split}
 &[\delta\hat{G}_{\mu}(a),\hat{x}_{\nu}]=i\frac{\partial \delta\hat{G}_{\mu}(a)}{\partial p^{\nu}} \\
 &=-[G_{\mu},\hat{x}_{\nu}]+i\eta_{\mu\nu}s(a\cdot G)+i(s+2)a_{\mu}G_{\nu}+i(s+1)a_{\nu}G_{\mu}, \\
\end{split}\end{equation}
which can be easily integrated over $p^{\nu}$. Before writing $ \hat{G}_{\mu}$ explicitly, it is convenient to find an operator that commutes with $\hat{x}_{\mu}$. We find an operator $\hat{y}_{\mu}$ so that
\begin{equation}
[\hat{x}_{\mu},\hat{y}_{\nu}]=0, \qquad \hat{y}_{\mu}=x_{\mu}+\gamma x_{\mu}(a\cdot p)+(\gamma-1)(x\cdot p)a_{\mu}+\beta(x\cdot a)p_{\mu},
\end{equation}
and define $f(\hat{y})$ as
\begin{equation}\label{f(y)}
f(\hat{y})=f(x)+\gamma(x\cdot \frac{\partial f}{\partial x})(a\cdot p)+(\gamma-1)(a\cdot \frac{\partial f}{\partial x})(x\cdot p)+\beta(a\cdot x)(\frac{\partial f}{\partial x}\cdot p),
\end{equation}
so that
\begin{equation}
[f(\hat{y}),\hat{x}_{\mu}]=0.
\end{equation}
 Finally we can write the operator for the force in the non-commutating space when $e=0$ as
 \begin{equation}\label{e=0force}
 \hat{F}_{\mu}(e=0)=G_{\mu}(\hat{y})+s(a\cdot G)p_{\mu}+(s+2)a_{\mu}(G\cdot p)+(s+1)G_{\mu}(a\cdot p),
 \end{equation}
 that is
 \begin{equation}\label{e=0force1}
 \hat{F}_{\mu}(e=0)=G_{\mu}(\hat{y})+ms(a\cdot G)\dot{x}_{\mu}+m(s+2)a_{\mu}(G\cdot \dot{x})+m(s+1)G_{\mu}(a\cdot \dot{x}).
 \end{equation}
From the above, we see that a neutral particle with mass $m$, moving in a $\kappa$-deformed Minkowski space-time can be interpreted as moving in an `electro-magnetic'-like background that couples proportionally to the deformation parameter $a_{\mu}$, because these corrections are linear in $\dot{x}$.
 
 \subsection{$e\neq 0$ case and corrections to the Lorentz force}
In Section 2 we have shown that the minimal coupling principle leads to well known Lorentz force, so we want to generalize the minimal coupling principle in a consistent way. If we postulate $\hat{\pi}_{\mu}=m\dot{\hat{x}}_{\mu}$ and the most simplest way to introduce gauge field $A_{\mu}$, as $\hat{\pi}_{\mu}=\hat{p}_{\mu}-eA_{\mu}(\hat{x} \;or\; \hat{y})$, then from the Jacobi identities and  
 \begin{equation}\label{anti2}
[\hat{\pi}_{\mu},\hat{x}_{\nu}]+[\hat{x}_{\mu},\hat{\pi}_{\nu}]=i(a_{\mu}\hat{\pi}_{\nu}-a_{\nu}\hat{\pi}_{\mu}),
\end{equation}
we get very restrictive conditions on $A_{\mu}$. It is better to understand minimal coupling principle as a way to introduce connection between canonical and kinetic momentum through a gauge field in a way that the commutation relations 
 $[\hat{\pi}_{\mu},\hat{x}_{\nu}]$ and $[\hat{p}_{\mu},\hat{x}_{\nu}]$ are the same by form, which is also true in the  commutating case. In correspondence with (\ref{comm-NC}) we write
 
 \begin{equation}\label{comm-NC1}
 [\hat{\pi}_{\mu},\hat{x}_{\nu}]=i\eta_{\mu\nu}(1+s(a\cdot \pi)) + i(s+2)a_{\mu}\pi_{\nu}+i(s+1)a_{\nu}\pi_{\mu}.
 \end{equation}
 We also write $\hat{\pi}_{\mu}$ as $\hat{\pi}_{\mu}=\pi_{\mu}+\delta\hat{\pi}_{\mu}(a)$ and from (\ref{comm-NC1}) we get explicitly
 \begin{equation}
 \hat{\pi}_{\mu}=\hat{p}_{\mu}-eA_{\mu}(\hat{y})-e[(s+2)(A\cdot p)a_{\mu}+s(A\cdot a)p_{\mu}+(s+1)A_{\mu}(a\cdot p)].
 \end{equation}
 Taking the derivative with respect to $\tau$ of (\ref{comm-NC1}) and using $\hat{F}_{\mu}=\frac{d \hat{\pi}_{\mu}}{d \tau}$, we get
 \begin{equation} [\hat{F}_{\mu},\hat{x}_{\nu}]+\frac{1}{m}[\hat{\pi}_{\mu},\hat{\pi}_{\mu}]=i\eta_{\mu\nu}s(a\cdot F)+i(s+2)a_{\mu}F_{\nu}+i(s+1)a_{\nu}F_{\mu}.
 \end{equation}
 Writing the force as $\hat{F}_{\mu}=F_{\mu}+\delta\hat{F}_{\mu}(a)$, we can get a equation for $\delta\hat{F}_{\mu}(a)$ as
 
 \begin{equation}\label{varF}
 [\delta\hat{F}_{\mu}(a),x_{\nu}]=[\hat{x}_{\nu}, F_{\mu}]-\frac{1}{m}[\hat{\pi}_{\mu},\hat{\pi}_{\nu}]+i\eta_{\mu\nu}s(a\cdot F)+i(s+2)a_{\mu}F_{\nu}+i(s+1)a_{\nu}F_{\mu}.
 \end{equation}
 where $F_{\mu}$ is the force from the commutative space and 
\begin{equation}\begin{split}\label{pi-pi}
[\hat{\pi}_{\mu},\hat{\pi}_{\nu}]&=-ie[F_{\mu\nu}+2(s+1)F_{\mu\nu}a\cdot p+i(\alpha+\beta)a\cdot \frac{\partial F_{\mu\nu}}{\partial x}+\gamma(x\cdot \frac{\partial F_{\mu\nu}}{\partial x})(a\cdot p)\\
&+(\gamma-1)(a\cdot\frac{\partial F_{\mu\nu}}{\partial x})(x\cdot p)
 + \beta(a\cdot x)(\frac{\partial F_{\mu\nu}}{\partial x}\cdot p)+sa^{\alpha}(F_{\alpha\nu}p_{\mu}-F_{\alpha\mu}p_{\nu})\\
 &+(s+2)(a_{\mu}F_{\alpha\nu}-a_{\nu}F_{\alpha\mu})p^{\alpha}+i\gamma(a_{\mu}\frac{\partial^{2}A_{\nu}}{\partial x_{\alpha}\partial x^{\alpha}}-a_{\nu}\frac{\partial^{2}A_{\mu}}{\partial x_{\alpha}\partial x^{\alpha}})]\\
 &+ie^{2}[(s+2)A^{\alpha}(a_{\mu}\frac{\partial A_{\nu}}{\partial x^{\alpha}}-a_{\nu}\frac{\partial A_{\mu}}{\partial x^{\alpha}})+s(a\cdot A)F_{\mu\nu}\\
 &+(s+1)a^{\alpha}(A_{\mu}\frac{\partial A_{\nu}}{\partial x^{\alpha}}-A_{\nu}\frac{\partial A_{\mu}}{\partial x^{\alpha}})-\frac{\partial A_{\nu}}{\partial x^{\alpha}}\frac{\partial A_{\mu}}{\partial x^{\beta}}(x^{\alpha}a^{\beta}-a^{\alpha}x^{\beta})].\\
\end{split}\end{equation}
 R.H.S. of equation (\ref{varF}) can be explicitly calculated and the L.H.S is  
 \begin{equation}
 [\delta\hat{F}_{\mu}(a),x_{\nu}]=i\frac{\partial(\delta\hat{F}_{\mu}(a))}{\partial p^{\nu}},
  \end{equation} 
  so we can integrate (\ref{varF})
over $p^{\nu}$. After tedious calculation and expressing everything in terms of $\dot{\hat{x}}_{\mu}$ we get
 \begin{equation} \hat{F}_{\mu}=\hat{G}_{\mu}+eF_{\mu\nu}(\hat{y})\dot{\hat{x}}^{\nu}+e\tilde{F}_{\mu\nu}\dot{\hat{x}}^{\nu}-m\Gamma_{\mu\nu\lambda}\dot{\hat{x}}^{\nu}\dot{\hat{x}}^{\lambda}+O(a\cdot e^{2})+O(a^{2}),
  \end{equation}
  where $\hat{G}_{\mu}=\hat{F}_{\mu}(e=0)$ is defined in (\ref{e=0force}), $F_{\mu\nu}(\hat{y})$ is defined like (\ref{f(y)}), that is
  \begin{equation}
F_{\mu\nu}(\hat{y})=F_{\mu\nu}(x)+\gamma(x\cdot \frac{\partial F_{\mu\nu}}{\partial x})(a\cdot p)+(\gamma-1)(a\cdot \frac{\partial F_{\mu\nu}}{\partial x})(x\cdot p)+\beta(a\cdot x)(\frac{\partial F_{\mu\nu}}{\partial x}\cdot p),
\end{equation}
  and the remaining two terms are 
 \begin{equation}\begin{split}
\tilde{F}_{\mu\nu}&=i[(\alpha+\beta)a\cdot \frac{\partial F_{\mu\nu}}{\partial x}
-\gamma(a_{\mu}\frac{\partial^{2}A_{\nu}}{\partial x_{\alpha}\partial x^{\alpha}}-a_{\nu}\frac{\partial^{2}A_{\mu}}{\partial x_{\alpha}\partial x^{\alpha}})], \\
\Gamma_{\mu\nu\lambda}&=e[(\alpha+\beta)F_{\mu\nu}a_{\lambda}+(\gamma+\beta)F_{\mu\rho}a^{\rho}\eta_{\lambda\nu}-sF_{\rho\nu}a^{\rho}\eta_{\mu\lambda}-(3\gamma-2\beta-1)F_{\mu\nu}a_{\lambda}].\\
\end{split}\end{equation}
Terms proportional to $\dot{x}$ can be interpreted as a correction to the Lorentz force due to the background electromagnetic field, and those proportional to $\dot{x}^{2}$ as quasi-gravitational effects caused by the background curvature induced by the non-commutativity of space-time. Both effects are proportional to $ea_{\mu}$.

 \subsection{$\kappa$-deformed Maxwell equations}
 In our approach all the Jacobi identities using $\hat{\pi}$, $\hat{p}$ and $\hat{x}$ are satisfied by construction up to the first order in the deformation parameter $a$. Formally the Jacobi identity
 \begin{equation}
[\hat{\pi}_{\mu},[\hat{\pi}_{\nu},\hat{\pi}_{\rho}]]+[\hat{\pi}_{\nu},[\hat{\pi}_{\rho},\hat{\pi}_{\mu}]]+[\hat{\pi}_{\rho},[\hat{\pi}_{\mu},\hat{\pi}_{\nu}]]=0,
\end{equation}
leads to 
\begin{equation}\label{kappahomo}
\partial_{\mu}\hat{F}_{\nu\rho}+\partial _{\nu}\hat{F}_{\rho\mu}+\partial _{\rho}\hat{F}_{\mu\nu}=i([\delta\hat{\pi}_{\mu}(a),F_{\nu\rho}]-e[A_{\mu},\delta\hat{F}_{\nu\rho}(a)]+cyclic(\mu,\nu,\rho)),
\end{equation}
where
\begin{equation}
[\hat{\pi}_{\mu},\hat{\pi}_{\nu}]\equiv-ie\hat{F}_{\mu\nu}=-ieF_{\mu\nu}(x)-ie\delta\hat{F}_{\mu\nu}(a),
\end{equation}
is given in (\ref{pi-pi}), so we see that $\hat{F}_{\mu\nu}$ is expressed in terms of operators from the commutative space. This is the $\kappa$-deformed analogue of the homogeneous Maxwell equation. The R.H.S of (\ref{kappahomo}) can be explicitly calculated in terms of commutative variables and fields $\vec{E}$ and $\vec{B}$, that satisfy the usual Maxwell equations. From 
\begin{equation}
[\hat{\pi}_{\nu},[\hat{\pi}_{\mu},[\hat{\pi}^{\mu},\hat{\pi}^{\nu}]]]=0,
\end{equation}
and by defining 
\begin{equation}
[\hat{\pi}_{\mu},[\hat{\pi}^{\mu},\hat{\pi}^{\nu}]]=e\hat{j}^{\nu},
\end{equation} 
we have 
\begin{equation}
[\hat{\pi}_{\mu},\hat{j}^{\mu}]=0,
\end{equation}
so that $\hat{j}$ is a conserved current and we formally have
\begin{equation}\label{kappainhomo}
\partial_{\mu}\hat{F}^{\mu\nu}=\hat{j}^{\nu}+i[\delta\hat{\pi}_{\mu} (a),F^{\mu\nu}]-ie[A_{\mu},\delta\hat{F}^{\mu\nu} (a)]+O(a^{2}).
\end{equation}
This is the $\kappa$-deformed analogue of the inhomogeneous Maxwell equation. R.H.S in (\ref{kappainhomo}) can also be explicitly calculated.

Now we study the $\kappa$-deformed space-time where $a_{\mu}=(a_{0},\vec{0})$, that is $a_{0}\equiv a=\kappa^{-1}$ and $a_{i}=0$. We define
\begin{equation}\begin{split}
\hat{F}_{0i}&=-\hat{E}_{i}, \qquad   F_{0i}=-E_{i},\\
  \hat{F}_{ij}&=-\epsilon_{ijk}\hat{B}_{k}, \qquad     F_{ij}=-\epsilon_{ijk}B_{k}.\\
\end{split}\end{equation}
 Note that $\hat{F}_{\mu\nu}$, $\hat{E}_{i}$ and $\hat{B}_{i}$  are functions of commutative operators $x$ and $p$ (see (\ref{pi-pi})) . Now we can rewrite the R.H.S of (\ref{kappahomo})  and (\ref{kappainhomo}) in terms of commutative electric and magnetic field and get
\begin{equation}
\vec{\nabla}\cdot\hat{\vec{B}}=a(\alpha+\beta)\dot{\vec{B}}\cdot\vec{p}   -ae(\vec{D}_{B}\cdot\vec{B}+s\vec{E}\cdot\vec{B}),
\end{equation}
\begin{equation}\begin{split}
\vec{\nabla}\times\hat{\vec{E}}+\frac{\partial \hat{\vec{B}}}{\partial t}=&-a[(\alpha+\beta)(\dot{\vec{B}}p_{0}-\dot{\vec{E}}\times\vec{p})+\gamma(p\cdot\frac{\partial \vec{B}}{\partial x}+\frac{\partial \vec{B}}{\partial x}\cdot p)]\\
&+ae[\vec{\Box}_{E} \times \vec{E}+D_{B}\vec{B}-(s+2)B_{i}\vec{\nabla}A_{i}],\\
\end{split}\end{equation}
\begin{equation}
\vec{\nabla}\cdot\hat{\vec{E}}=\hat{\rho}-a(\alpha+\beta)(p_{0}\vec{\nabla}\cdot\vec{E}-\dot{\vec{E}}\cdot\vec{p})
+ae(\vec{D}_{E}\cdot\vec{E}-(s+2)\vec{B}^{2}),
\end{equation}
\begin{equation}\begin{split}
\vec{\nabla}\times\hat{\vec{B}}-\frac{\partial \hat{\vec{E}}}{\partial t}&=\hat{\vec{j}}+a[(\alpha+\beta)(p_{0}\dot{\vec{E}}+\dot{\vec{E}}p_{0}-p_{0}\vec{\nabla}\times\vec{B}-\dot{\vec{B}}\times\vec{p})+\gamma(p\cdot\frac{\partial\vec{E}}{\partial x}+\frac{\partial\vec{E}}{\partial x}\cdot p)]\\
&-ae(\vec{\Box}_{B}\times\vec{B}+D_{E}\vec{E}+(s+2)\vec{B}\times\vec{E}+sE_{i}\vec{\nabla}A_{i}).\\
\end{split}\end{equation}
These equations represent the $\kappa$-deformed set of Maxwell equations. The operators $\vec{D}_{B}$, $\vec{D}_{E}$, $D_{B}$, $D_{E}$, $\vec{\Box}_{B}\times$ and $\vec{\Box}_{E}\times$ are given as follows
\begin{equation}\begin{split}\label{operators}
&D_{B}=(\vec{r}\cdot\vec{\nabla}) \phi\frac{\partial}{\partial t}-\dot{\phi}(\vec{r}\cdot\vec{\nabla})+(2s+3)\phi\frac{\partial}{\partial t}-2(s+1)\dot{\phi}+(s+2)(\vec{A}\cdot\vec{\nabla})+(s+2)(\vec{\nabla}\cdot\vec{A}),\\
&D_{E}=D_{B}+s\phi\frac{\partial}{\partial t}-2(s+1)\dot{\phi}-2(s+1)(\vec{\nabla}\cdot\vec{A}),\\
&\vec{D}_{B}=(\vec{r}\cdot\vec{\nabla})\vec{A}\frac{\partial}{\partial t}-\dot{\vec{A}}(\vec{r}\cdot\vec{\nabla})+(s+1)\vec{A}\frac{\partial}{\partial t}-2(s+1)\dot{\vec{A}},\\
&\vec{D}_{E}=-\vec{D}_{B}+s\phi\vec{\nabla}+2(s+1)\dot{\vec{A}},\\
&\vec{\Box}_{B}\times=(\vec{r}\cdot\vec{\nabla})\vec{A}\times\frac{\partial}{\partial t}-\dot{\vec{A}}\times(\vec{r}\cdot\vec{\nabla})
-s\phi\vec{\nabla}\times+(s+1)\vec{A}\times\frac{\partial}{\partial t}-(3s+4)\dot{\vec{A}}\times,\\
&\vec{\Box}_{E}\times=-\vec{\Box}_{B}\times-s\phi\vec{\nabla}\times.\\
\end{split}\end{equation}

\subsection{$Natural$ realization}
We see that all the corrections  to commutative electrodynamics depend on the realization of operator $\hat{x}_{\mu}$ 
\begin{equation}\label{hatx}
\hat{x}_{\mu}=x_{\mu}+\alpha x_{\mu}(a\cdot p)+\beta(x\cdot a)p_{\mu}+\gamma(x\cdot p)a_{\mu},
\end{equation}
that is on the parameters $\alpha$, $\beta$ and $\gamma$. We are going to investigate the so called $natural$ realization \cite{natural}. The most easiest way to get the $natural$ realization is to demand that (\ref{hatx}) is hermitian, that is $\hat{x}^{\dagger}=\hat{x}$ and put  $\gamma=0$, then we get $\alpha=-1$, and $\beta=1$, and for operator $\hat{x}$ in $natural$ realization we have
\begin{equation}
\hat{x}_{\mu}^{nat}=x_{\mu}[1-(a\cdot p)]+(x\cdot a)p_{\mu}.
\end{equation}
The parameter $s$ becomes $s^{nat}=-1$. For the complex operators given in (\ref{operators})  in $natural$ realization we get
\begin{equation}\begin{split}
\vec{D}_{B}&=(\vec{r}\cdot\vec{\nabla})\vec{A}\frac{\partial}{\partial t}-\dot{\vec{A}}(\vec{r}\cdot\vec{\nabla}),\\
\vec{D}_{E}&=-\vec{D}_{B}-\phi\vec{\nabla},\\
D_{B}&=(\vec{r}\cdot\vec{\nabla})\phi\frac{\partial}{\partial t}-\dot{\phi}(\vec{r}\cdot\vec{\nabla})+\phi\frac{\partial}{\partial t}+(\vec{A}\cdot\vec{\nabla})+(\vec{\nabla}\cdot\vec{A}),\\
D_{E}&=D_{B}-\phi\frac{\partial}{\partial t},\\
\vec{\Box}_{E}\times&=\dot{\vec{A}}(\vec{r}\cdot{\vec{\nabla}})-(\vec{r}\cdot\vec{\nabla})\vec{A}\times\frac{\partial}{\partial t}+\dot{\vec{A}}\times,\\
\vec{\Box}_{B}\times&=-\vec{\Box}_{E}\times + \phi\vec{\nabla}\times.\\
\end{split}\end{equation}
And for the $\kappa$-deformed Maxwell equations in the $natural$ realization we finally have 
\begin{equation}\begin{split}
\vec{\nabla}\cdot\hat{\vec{B}}&=-ae(\vec{D}_{B}\cdot\vec{B}-\vec{E}\cdot\vec{B}),\\
\vec{\nabla}\times\hat{\vec{E}}+\frac{\partial \hat{\vec{B}}}{\partial t}&=ae(\vec{\Box}_{E} \times \vec{E}+D_{B}\vec{B}-B_{i}\vec{\nabla}A_{i}),\\
\vec{\nabla}\cdot\hat{\vec{E}}&=\hat{\rho}+ae(\vec{D}_{E}\cdot\vec{E}-\vec{B}^{2}),\\
\vec{\nabla}\times\hat{\vec{B}}-\frac{\partial \hat{\vec{E}}}{\partial t}&=\hat{\vec{j}}-ae(\vec{\Box}_{B}\times\vec{B}+D_{E}\vec{E}+\vec{B}\times\vec{E}-E_{i}\vec{\nabla}A_{i}).\\
\end{split}\end{equation}

For the the force operator we have
\begin{equation}\label{lf}
\hat{F}_{\mu}^{nat}=\hat{G}_{\mu}^{nat}+eF_{\mu\nu}^{nat}(\hat{y})\dot{\hat{x}}^{\nu}-m\Gamma_{\mu\nu\lambda}^{nat}\dot{x}^{\nu}\dot{x}^{\lambda},
\end{equation}
where
\begin{equation}\begin{split}
\tilde{F}_{\mu\nu}^{nat}=& 0,\\
\Gamma_{\mu\nu\lambda}^{nat}=& ae(-F_{\mu0}\eta_{\lambda\nu}+F_{0\nu}\eta_{\mu\lambda}+3F_{\mu\nu}\delta_{\lambda}^{0}),\\
F_{\mu\nu}^{nat}(\hat{y})=& F_{\mu\nu}+at(\frac{\partial F_{\mu\nu}}{\partial x}\cdot p)-a\dot{F}_{\mu\nu}(x\cdot p),\\
\hat{G}_{\mu}^{nat}=& G_{\mu}(\hat{y})-amG_{0}\dot{x}_{\mu}+am(G\cdot\dot{x})\delta_{\mu}^{0}.\\
\end{split}\end{equation}
Note (see Eqn.{\ref{lf}) that the force depends not only on the charge of the particle, but on its mass also. This mass depends vanishes in the limit of $a\to 0$.


\section{Conclusion}

In this paper, we have constructed force equation and Maxwell's equation on $\kappa$-deformed space-time. For this construction, we have generalized a variation of Feynman's approach \cite{minfeyn} to $\kappa$-deformed non-commutative space-time. This approach also starts with the same assumptions as in Feynman's approach \cite{Dyson, Tanimura}, and used the notion of canonical conjugate momenta and their commutators (or Poisson brackets) with coordinates. Then as in the Feynman's approach, by repeated use of Jacobi identity, force equation and Maxwell's equations are derived. The main differences in this approach are the use of minimal coupling prescription for the gauge field, and the existence of a classical limit. This approach \cite{minfeyn, Dyson, Tanimura} allows us  to take the classical limit which is obtained by replacing $(i\hbar)^{-1}[~]$ with $\{~\}_{PB}$. 

We have obtained the $\kappa$-dependent modification to the Newtons force equation in subsection. 3.1. Then, we introduce the gauge field in subsection 3.2, and derive the Lorentz force equation as well as Maxwell's equations, in the $\kappa$-space-time. The additional contributions due to $\kappa$-deformation of space-time to the force equation that are linear in $\dot{x}$ can be interpreted as due to a background electromagnetic field and those proportional to $\dot{x}^{2}$ as a induced curvature of space-time. This is in similar spirit as the induced gravity in Moyal space-time considered in \cite{vor}. Here, these corrections are obtained up to first order in the deformation parameter. This change in the Lorentz force equation will affect the trajectories of charged particles in external electromagnetic fields. This can lead to possible, observable effects in the beams of high energy accelerators. It is clear that this effects would violate Lorentz symmetry and modify the dispersion relations. This aspects are in detail investigated in \cite{ab}.

The $\kappa$-dependent corrections to force equation and Maxwell's equations changes with the choice of realization of non-commutative coordinates we use. We have investigated this modification for natural realization \cite{natural} in subsection. 3.4. This realization is hermitian and has the property that the corresponding momentum transforms as a $4$-vector under Lorentz algebra.

The $\kappa$-deformed  Maxwell equations obtained here are  complicated, even in the natural realization. With further simplifying assumptions, they can be compared easily with what we know in the commutative case. For the static case, with  $\hat{\rho}, E,$ and $\phi$ set to zero we get the equations for $\kappa$-deformed magnetostatic. They are 
\begin{equation}\begin{split}
\vec{\nabla}\cdot\hat{\vec{B}}&=0\\
\vec{\nabla}\times\hat{\vec{B}}&=\hat{\vec{j}}\\
\end{split}\end{equation}
In electrostatic limit there are additional terms. The framework employed here to derive the force equation and Maxwell's equations, allow us to replace the (quantum) commutators with corresponding Poisson brackets to get the classical result \cite{minfeyn, Dyson, Tanimura}. Thus by substituting commutator with Poisson bracket, that is $\frac{1}{i\hbar}[,]\rightarrow \left\{,\right\}$  and all the operators go to $c$-number functions. Now the parameter $\tau$ becomes proper time of a particle.

In this classical limit we calculate the corrections to the classical Coulomb force between two test particles of charge $e$ at rest separated at a distance $r$ in the $\kappa$-deformed non-commutative space time. Since, we consider non relativistic case, we get $\frac{d \tau}{d t}=1$. By setting $\dot{x}_{i}=0$ and with vanishing $G_{\mu},B,A$ and $E=\frac{e}{4\pi r^{2}}$ in (\ref{lf}), we get
\begin{equation}
\hat{F}=\frac{e^{2}}{4\pi r^{2}}(1-2am)
\end{equation}
Thus we see that the Coulomb law does not change the form.  The effect of non-commutativity can be interpreted as change in the charge of the particle $e \rightarrow {e(1-2am)}^{\frac{1}{2}}$. This shows that the electrodynamics  depends on the 
mass of the particle as well as its charge. Same feature was shown in \cite{eh} also.

Next, we consider the case of a particle of mass $m$ and charge $e$, moving in a constant external electric field $E$, with a velocity ${\vec v}$. From Eqn.(\ref{lf}), we find 
\begin{equation}\begin{split}
\hat{\vec{F}} &=e\vec{E}(\gamma-am(2\gamma^{2}+\vec{v}^{2}))-aem(\vec{E}\cdot\vec{v})\vec{v} \\
&=\gamma e\vec{E}-ame\vec{E}(2\gamma^{2}+\vec{v}^{2})-aem(\vec{E}\cdot\vec{v})\vec{v}, \\
\end{split}\end{equation}
where $\gamma=(1-\vec{v}^{2})^{1/2}$. With further choice ${\vec E}=(E, 0, 0)$, we get 
\begin{equation}\label{conelf}
\begin{split}
\hat{{F}}^{x} &=eE(\gamma-am(2\gamma^{2}+\vec{v}^{2}))-aem(Ev^{x})v^{x} \\
\hat{{F}}^{y} &=-aem(Ev^{x})v^{y} \\
\hat{{F}}^{z} &=-aem(Ev^{x})v^{z}. \\
\end{split}\end{equation}
These can be solved to get 
\begin{equation}\begin{split}
\dot{\hat{y}}(\tau)&=\dot{y}(0)e^{-aeEx(\tau)} \\
\hat{y}(\tau)&=\dot{y}(0)\tau-\dot{y}(0)aeE\int d\tau x(\tau)+y(0).
\end{split}\end{equation}
It is easy to see that  $\hat{z}(\tau)$ also obeys the same  equation as $\hat{y}(\tau)$, showing the $a$ dependent modification of $\hat{y}(\tau)$ and $\hat{z}(\tau)$ as deviations from classical trajectories. This pure noncomutative effect may also put some bounds on the parameter $a$. With initial conditions, $\dot{y}(0)=\dot{z}(0)=0$,
we get $a$ dependent modified equation
\begin{equation}
\ddot{\hat{x}}=eE/m\frac{1}{\sqrt{1-\dot{x}^{2}}}-\frac{2aeE}{1-\dot{x}^{2}}-2aeE\dot{x}^{2}.
\end{equation}
Eqn. (\ref{conelf}) shows that the $a$ dependent modification to force equation also depends on the mass of the particle apart from its charge.

We also note that, to the first order in the deformation parameter, there is no corrections to the Newtons law of gravity. This can be seen by setting  $\dot{x}_{i}, G_{0}=0$ and $G_{i}=-G\frac{m^{2}}{r^{2}}$  in Eqn. (34). This is different from what was shown in \cite{akk}. In \cite{akk}, using the Hamiltonian framework, modification to Newton's  second law due to $\kappa$-deformation was derived. The (non-relativistic) Hamiltonian for a free particle was obtained  by taking the appropriate limit of the energy-momentum relation valid in $\kappa$-space-time. Here, up to first order in the deformation parameter $a$, the effect of deformation was to modify the mass $m\to m(1+am)$. Though the $\frac{1}{r}$ potential was also modified (up to first order in $a$), this term did not contribute to the force equation. Here also we do see the same feature.

Also, the fact that the force equation was obtained in \cite{akk} using the Hamiltonian framework different from what we get here raises the question whether the equations (of motion) obtained here are derivable from a Hamiltonian or a Lagrangian. In the commutative case, the condition for existence of Lagrangian/ Hamiltonian from which equations of motion can be derived had been studied \cite{AH,RJH}.This problem, in the Moyal space was investigated in \cite{cg}. We plan to study this in the case of $\kappa$ space-time.

In the Feynman's approach, due to the non-vanishing commutators between the coordinates and velocities, the rotation symmetry is broken and it was shown that by including magnetic angular momentum, this symmetry can be restored  \cite{abyg1}. Inclusion of magnetic monopoles in the Feynman's approach was also considered.  We plan to address these issues separately. Generalizing the method adopted here for general relativistic case, where the metric will depend on the space-time coordinate is of immense interest. This work is in progress and will be reported elsewhere.

\noindent{\bf Acknowledgment}\\
We  thank K. S. Gupta for useful discussions. TJ and SM were supported by
the Ministry of Science and Technology of the Republic of Croatia under contract No.
098-0000000-2865.

\end{document}